\documentclass[10pt,final,doublecolumn]{IEEEtran}
\usepackage{amsmath}
\usepackage{latexsym}
\usepackage{graphicx}
\usepackage{bbding}
\usepackage{enumerate}
\usepackage{multicol}
\usepackage{color}
\usepackage{slashbox}
\IEEEoverridecommandlockouts
\begin{document}

\title{Wireless Energy and Information Transfer Tradeoff for Limited Feedback Multi-Antenna Systems with Energy Beamforming}
\author{\normalsize
Xiaoming~Chen, Chau~Yuen, and Zhaoyang~Zhang
\thanks{Copyright (c) 2013 IEEE. Personal use of this material
is permitted. However, permission to use this material for any
other purposes must be obtained from the IEEE by sending a request
to pubs-permissions@ieee.org.}
\thanks{This work was supported by the Natural
Science Foundation Program of China (No. 61300195), the Natural
Science Foundation Program of Jiangsu Province (No. 2013001012),
the open research fund of State Key Laboratory of Integrated
Services Networks, Xidian University (No. ISN14-01), the Doctoral
Fund of Ministry of Education of China (No. 20123218120022) and
the Singapore University Technology and Design (No.
SUTD-ZJU/RES/02/2011).}
\thanks{Xiaoming~Chen ({\tt chenxiaoming@nuaa.edu.cn}) is with the
College of Electronic and Information Engineering, Nanjing
University of Aeronautics and Astronautics, and also with the
State Key Laboratory of Integrated Services Networks, Xidian
University, China. Chau~Yuen ({\tt yuenchau@sutd.edu.sg}) is with
the Singapore University of Technology and Design, Singapore.
Zhaoyang~Zhang ({\tt ning\_ming@zju.edu.cn}) is with the
Department of Information Science and Electronic Engineering,
Zhejiang University, Hangzhou 310027, China.}}\maketitle

\begin{abstract}
In this paper, we consider a multi-antenna system where the
receiver should harvest energy from the transmitter by wireless
energy transfer to support its wireless information transmission.
In order to maximize the harvesting energy, we propose to perform
adaptive \emph{energy beamforming} according to the instantaneous
channel state information (CSI). To help the transmitter to obtain
the CSI for energy beamforming, we further propose a win-win CSI
quantization feedback strategy, so as to improve the efficiencies
of both power and information transmission. The focus of this
paper is on the tradeoff of wireless energy and information
transfer by adjusting the transfer duration with a total duration
constraint. Through revealing the relationship between transmit
power, transfer duration and feedback amount, we derive two
wireless energy and information transfer tradeoff schemes by
maximizing an upper bound and an approximate lower bound of the
average information transmission rate, respectively. Moreover, the
impact of imperfect CSI at the receiver is investigated and the
corresponding wireless energy and information transfer tradeoff
scheme is also given. Finally, numerical results validate the
effectiveness of the proposed schemes.
\end{abstract}


\section{Introduction}
Recently, wireless energy transfer has aroused general interests
in wireless research community, as it can effectively prolong the
lifetime of the power-limited node or network in a relative simple
way [1]-[3]. As a typical example, in medical area, the implanted
equipment in body can be powered through wireless power transfer
\cite{Medical}.

In general, wireless energy transfer from a power source to a
receiver is implemented through electromagnetic propagation
\cite{WET4}. Since electronic energy is propagated isotropically
if the transmit antenna is isotropic, the receiver only harvests a
portion of the transmitted energy without specific control,
resulting in a low energy transfer efficiency. In order to
maximize the harvested energy, it is necessary to coordinate
transmit direction to the receiver, namely energy beamforming. The
key of the energy beamforming is the achievement of channel state
information (CSI) at the power source. Inspired by signal
beamforming in multi-antenna systems \cite{Beamforming1}
\cite{Beamforming2}, we also propose to use quantization codebooks
of limited size to convey the CSI from the receiver to the power
source. Based on the feedback CSI, the power source performs
adaptive energy beamforming.

The ultimate goal of wireless energy transfer is to fulfill the
need of the receiver for work. As an example, the implanted
equipment transmits the medical data to the outside receiver with
the harvested energy. However, most of previous analogous works
analyze and design the wireless energy transfer without
considering the utilization of the harvested energy. There are
some works joint considering wireless energy and information
transfer in the point-to-point system \cite{WEIT1} and the
multiuser downlink \cite{WEIT2}, but the harvested energy is not
used for information energy, they only share a common transmitter.
In this paper, we consider joint wireless energy and information
transfer in a multi-antenna system employing energy beamforming,
but the power receiver also plays a role as the information
transmitter. Specifically, the power receiver, such as the
implanted equipment, transmits the information with the harvested
energy. For such a system, since energy harvesting and information
transmission are impossibly carried out simultaneously, the time
slot should be divided into the harvesting and the transmitting
components. The larger harvesting duration denotes higher
available transmit power, but also shortens the transmitting
duration, which may lead to a lower average transmission rate.
Thus, for a given duration of one time slot, it is imperative to
determine the optimal time allocation between energy and
information transfer, namely the wireless energy and information
transfer tradeoff. The focus of this paper is on the optimal
tradeoff with the goal of maximizing the average information
transmission rate.

As mentioned above, energy beamforming has a great impact on
wireless power transfer, and thus the ultimate average information
transmission rate. On the other hand, the amount of CSI at the
power source has a tight connection to the efficiency of power
transfer. Therefore, we give a detailed investigation of the
impact of the feedback amount on the average information
transmission rate and derive the tradeoff between the durations of
energy and information transfer in terms of feedback amount from
the perspectives of maximizing an upper bound and an approximate
lower bound on the average information transmission rate,
respectively. Moreover, considering the limited channel estimation
capacity, there will be estimation error at the power receiver.
Under such a condition, we also analyze the impact of estimation
error on the average information transmission rate, and obtain the
corresponding wireless energy and information transfer tradeoff.

The rest of this paper is organized as follows: Section II gives a
brief introduction of the considered wireless energy and
information transfer protocol for a limited feedback multi-antenna
system with energy beamforming. Section III focuses on the
analysis and design of the wireless energy and information
transfer tradeoff with limited feedback. The impact of imperfect
CSI caused by channel estimation error is investigated in Section
IV. Section V presents several numerical results to validate the
performance of the proposed algorithms, and finally Section VI
concludes the whole paper.

\section{System Model}

\begin{figure}[h] \centering
\includegraphics [width=0.45\textwidth] {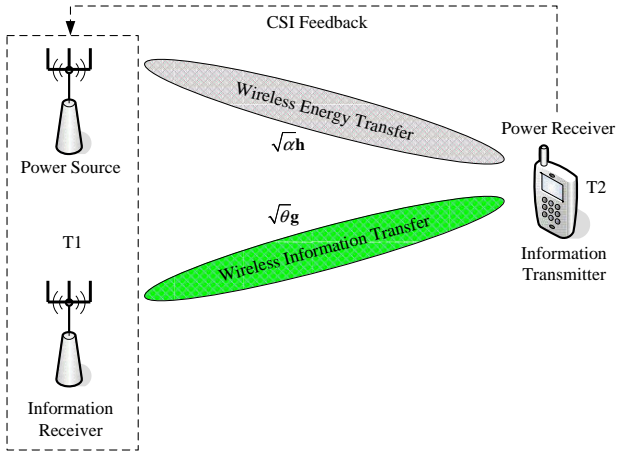}
\caption {A model of the multi-antenna system employing wireless
energy and information transfer.} \label{Fig1}
\end{figure}

We consider a frequency division duplex (FDD) multi-antenna system
employing wireless energy and information transfer, as shown in
Fig.\ref{Fig1}, where the power source and the information
receiver are equipped with $N_t$ antennas to enhance the energy
transfer efficiency by energy beamforming and improve the average
information transmission rate by receive combining, respectively.
In fact, the power source and the information receiver can be
integrated into one equipment T$_1$ and the work mode is switched
according to the proposed wireless energy and information transfer
tradeoff scheme. Similarly, T$_2$ can play the roles as both the
power receiver and the information transmitter. Due to space
limitation, T$_2$ deploys a single antenna for energy harvesting
and information transmitting. T$_2$ only has limited power to
maintain the active state, it needs to harvest the enough energy
for information transmission from the outside, such as T$_1$.
Considering the limited storage, T$_2$ should be charged from
T$_1$ slot by slot. We use $\sqrt{\alpha}\textbf{h}$ to denote the
$N_t$ dimensional channel vector from T$_1$ to T$_2$, where
$\alpha$ represents the path loss and $\textbf{h}$ denotes the
channel fast fading component that is an $N_t$ dimensional
zero-mean complex Gaussian vector having a real part with variance
1 and an imaginary part with variance 1. According to the law of
energy conservation, the harvesting power at T$_2$ from T$_1$ can
be expressed as \cite{WEIT2}
\begin{equation}
P_{harv}=\eta\alpha P_1|\textbf{h}^H\textbf{w}|^2, \label{eqn1}
\end{equation}
where $P_1$ is the transmit power of T$_1$, the constant parameter
$0\leq\eta\leq1$ is the efficiency ratio at T$_2$ for converting
the harvested energy to the electrical energy to be stored.
Following \cite{Conversionefficiency}, we assume $\eta=0.8$ in
this paper. $\textbf{w}$, as the energy beamforming vector with
unit norm, is used to adaptively adjust the energy transmit
direction according to the instantaneous channel state
$\textbf{h}$, so as to maximize the harvesting power at T$_2$.
Clearly, the CSI at T$_1$ has a great effect on the performance of
energy transfer. As a simple example, if full CSI is available,
$\textbf{w}=\textbf{h}/\|\textbf{h}\|$, namely maximum ratio
transmission (MRT), can achieve the optimal performance. However,
as T$_1$ is at the transmit side of the channel
$\sqrt{\alpha}\textbf{h}$, it is difficult to obtain the CSI by
itself. Inspired by signal beamforming in multi-antenna systems,
we propose to use a quantization codebook
$\mathcal{W}=\{\textbf{w}_1,\cdots,\textbf{w}_{2^B}\}$ of size
$2^{B}$ to convey the optimal beam from T$_2$ to T$_1$ for each
time slot. The quantization codebook is designed according to the
distribution of the channel direction vector
$\textbf{h}/\|\textbf{h}\|$ by making use of the vector
quantization (VQ) method and stored at both T$_1$ and T$_2$ in
advance. T$_2$ selects the optimal quantization codeword, namely
the beam, so as to maximize the effective channel gain.
Mathematically, the codeword selection criterion can be expressed
as
\begin{equation}
i^{\star}=\arg\max\limits_{\textbf{w}_i\in\mathcal{W}}|\textbf{h}^H\textbf{w}_i|^2.
\label{eqn2}
\end{equation}
T$_2$ conveys the optimal codeword index $i^{\star}$ to T$_1$, and
T$_1$ recovers the beam from the same codebook, then uses it for
energy beamforming.

\begin{figure}[h] \centering
\includegraphics [width=0.45\textwidth] {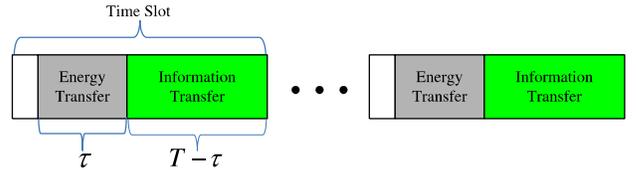}
\caption {Wireless energy and information transfer protocol}
\label{Fig2}
\end{figure}

The whole system is operated in slotted time of length $T$, as
shown in Fig.\ref{Fig2}. At the beginning of each time slot,
$i^{\star}$ is fed back from T$_2$ to T$_1$. It is assumed that
the energy and duration for feedback can be negligible due to the
small amount of transmission. Then, T$_1$ performs wireless energy
transfer by energy beamforming with the duration $\tau$, and the
harvesting energy at T$_2$ is given by
\begin{equation}
Q_{harv}=\eta\alpha
P_1|\textbf{h}^H\textbf{w}_{i^{\star}}|^2\tau.\label{eqn3}
\end{equation}
After that, T$_1$ and T$_2$ toggle to the wireless information
transfer mode. With the harvesting energy $Q_{harv}$, T$_2$
transmits the information to T$_1$ in the rest of the time slot
$T-\tau$, and the receive signal can be expressed as
\begin{equation}
\textbf{y}=\sqrt{\frac{Q_{harv}}{T-\tau}\theta}\textbf{g}s+\textbf{n},\label{eqn4}
\end{equation}
where $s$ is the normalized transmit signal, $\textbf{y}$ is the
$N_t$ dimensional receive signal vector, and $\textbf{n}$ is the
additive Gaussian white noise with zero mean and variance matrix
$\sigma^2\textbf{I}_{N_t}$. $\sqrt{\theta}\textbf{g}$ denotes the
channel from T$_2$ to T$_1$, where $\theta$ is the path loss and
$\textbf{g}$ is the channel fast fading vector, which is
independent of $\textbf{h}$. $\frac{Q_{harv}}{T-\tau}$ is the
transmit power. Assuming perfect CSI at T$_1$, maximum ratio
combining (MRC) is performed to maximize the average information
transmission rate, which is given by
\begin{eqnarray}
&&\!\!\!\!R(\tau)\nonumber\\
&=&\!\!\!\!E\left[\frac{T-\tau}{T}\log_2\left(1+\frac{Q_{harv}\theta\|\textbf{g}\|^2}{(T-\tau)\sigma^2}\right)\right]\nonumber\\
&=&\!\!\!\!\frac{T-\tau}{T}E\left[\log_2\left(1+\frac{\eta\alpha\theta
P_1\|\textbf{h}\|^2|\tilde{\textbf{h}}^H\textbf{w}_{i^{\star}}|^2\|\textbf{g}\|^2\tau}{(T-\tau)\sigma^2}\right)\right],\label{eqn5}
\end{eqnarray}
where $\tilde{\textbf{h}}=\textbf{h}/\|\textbf{h}\|$ is the
channel direction vector, and $E[\cdot]$ denotes the expectation
operation.

\section{Wireless Energy and Information Transfer Tradeoff}
In this section, we focus on the wireless energy and information
tradeoff to optimize the average information transmission rate in
(\ref{eqn5}). Specifically, we attempt to derive the optimal
energy transfer duration $\tau$ so as to maximize the average
information transmission rate. Due to multiple random variables in
(\ref{eqn5}), it is difficult to obtain the closed-form expression
of the average information transmission rate. For tractability, we
turn to derive the tradeoff from the perspectives of the
maximization of an upper and lower bounds of the average
information transmission rate, respectively.

\subsection{The Upper Bound Case}
Since $\log_2(1+x)$ is a concave function with respect to $x$,
according to Jensen's inequality, we have
\begin{eqnarray}
&&\!\!\!\!R(\tau)\nonumber\\
&\leq&\!\!\!\!\frac{T-\tau}{T}\log_2\left(1+\frac{\eta\alpha\theta
P_1E\left[\|\textbf{h}\|^2|\tilde{\textbf{h}}^H\textbf{w}_{i^{\star}}|^2\|\textbf{g}\|^2\right]\tau}{(T-\tau)\sigma^2}\right)\nonumber\\
&=&\!\!\!\!\frac{T-\tau}{T}\log_2\left(1+\frac{aE\left[\|\textbf{h}\|^2\right]
E\left[|\tilde{\textbf{h}}^H\textbf{w}_{i^{\star}}|^2\right]E\left[\|\textbf{g}\|^2\right]\tau}{T-\tau}\right),\nonumber\\\label{eqn6}
\end{eqnarray}
where $a=\eta\alpha\theta P_1/\sigma^2$, and (\ref{eqn6}) holds
true since $\textbf{h}$ and $\textbf{g}$ are independent of each
other in such a FDD system by assuming the frequency gap is larger
than the correlation bandwidth. Considering both the real and
imaginary parts of $\textbf{h}$ and $\textbf{g}$ are distributed
according to $\mathcal{CN}(0,\textbf{I}_{N_t})$, so we have
$E\left[\|\textbf{h}\|^2\right]=E\left[\|\textbf{g}\|^2\right]=2N_t$.
Moreover, for the squared inner product of two $N_t$ dimensional
vectors uniformly distributed on the unit sphere, it is $\beta(1,
N_t-1)$ distributed. Thus,
$|\tilde{\textbf{h}}^H\textbf{w}_{i^{\star}}|^2$ is the maximum of
$2^B$ i.i.d. $\beta(1, N_t-1)$ distributed random variables caused
by the beam selection in (\ref{eqn1}), and its probability density
function (pdf) can be expressed as
\begin{eqnarray}
f(x)=2^{B}(N_t-1)(1-x)^{N_t-2}\bigg(1-(1-x)^{N_t-1}\bigg)^{2^{B}-1}.\label{eqn7}
\end{eqnarray}
Then, the expectation of
$|\tilde{\textbf{h}}^H\textbf{w}_{i^{\star}}|^2$ can be computed
as
\begin{eqnarray}
E\bigg[\left|\tilde{\textbf{h}}_l^H\textbf{w}_{i^{\star}}\right|^2\bigg]&=&\int_0^{\infty}xf(x)dx\nonumber\\
&=&1-2^{B}Beta\left(2^{B},\frac{N_t}{N_t-1}\right).\label{eqn8}
\end{eqnarray}
where $Beta(a,b)$ denotes the beta function. By utilizing the
property of beta function, we have the following relationship
\cite{RVQ}
\begin{equation}
1-2^{-\frac{B}{N_t-1}}<
E\bigg[\left|\tilde{\textbf{h}}_l^H\textbf{w}_{i^{\star}}\right|^2\bigg]<1-\frac{N_t-1}{N_t}2^{-\frac{B}{N_t-1}}.\label{eqn34}
\end{equation}
Substituting (\ref{eqn34}) into (\ref{eqn6}), an upper bound on
the average information transmission rate can be expressed as
\begin{equation}
R^u(\tau)=\frac{T-\tau}{T}\log_2\left(1+\frac{4aN_t^2\left(1-\frac{N_t-1}{N_t}2^{-\frac{B}{N_t-1}}\right)\tau}{T-\tau}\right).\label{eqn9}
\end{equation}
Hence, the wireless energy and information transfer tradeoff based
on the upper bound in (\ref{eqn9}) is equivalent to the following
optimization problem
\begin{eqnarray}
J_1\!\!\!\!\!&:&\!\!\!\!\!
\max\limits_{\tau}\frac{T-\tau}{T}\log_2\left(1+\frac{b\tau}{T-\tau}\right)\label{eqn10}\\
&s.t.&0\leq\tau\leq T,\label{eqn11}
\end{eqnarray}
where
$b=4aN_t^2\left(1-\frac{N_t-1}{N_t}2^{-\frac{B}{N_t-1}}\right)$.
$J_1$ is an optimization problem with a single continuous
variable. Taking the derivative of (\ref{eqn10}) with respect to
$\tau$, and letting the derivative equal to zero, we have
\begin{equation}
\frac{b}{(T+(b-1)\tau)\ln2}-\frac{1}{T}\log_2\left(1+\frac{b\tau}{T-\tau}\right)=0.\label{eqn12}
\end{equation}
Let $\tau_1=0$, $\tau_2=T$ and $\tau_3$ be the solution of
equality (\ref{eqn12}), then the optimal $\tau$ can be determined
according to the following criterion:
\begin{equation}
\tau^{\star}=\arg\max\limits_{\tau\in\Omega_1}R^{u}(\tau),\label{eqn13}
\end{equation}
where $\Omega_1=\{\tau_1,\tau_2,\tau_3\}$.

Thereby, we achieve the wireless energy and information transfer
tradeoff by maximizing an upper bound of the average information
transmission rate.

\subsection{The Lower Bound Case}
For the average information transmission rate in (\ref{eqn5}), by
some mathematical operations, it can be approximated as
\begin{eqnarray}
R(\tau)
&=&\frac{T-\tau}{T}E\bigg[\log_2\bigg(1\nonumber\\
&+&\exp\left(\ln\left(\frac{\eta\alpha\theta
P_1\|\textbf{h}\|^2|\tilde{\textbf{h}}^H\textbf{w}_{i^{\star}}|^2\|\textbf{g}\|^2\tau}{(T-\tau)\sigma^2}\right)\right)\bigg)\bigg]\nonumber
\end{eqnarray}
\begin{eqnarray}
&\geq&\frac{T-\tau}{T}\log_2\bigg(1\nonumber\\
&+&\exp\left(E\left[\ln\left({a\|\textbf{h}\|^2|\tilde{\textbf{h}}^H\textbf{w}_{i^{\star}}|^2\|\textbf{g}\|^2\frac{\tau}{T-\tau}}\right)\right]\right)\bigg)\label{eqn14}\\
&\approx&\frac{T-\tau}{T}\log_2\bigg(1\nonumber\\
&+&\exp\left(E\left[\ln\left({a\|\textbf{h}\|^2\|\textbf{g}\|^2}\left(1-2^{-\frac{B}{N_t-1}}\right)\frac{\tau}{T-\tau}\right)\right]\right)\bigg)\nonumber\\\label{eqn15}\\
&=&\frac{T-\tau}{T}\log_2\left(1+\frac{c\tau}{T-\tau}\exp\left(E\left[\ln\left(\|\textbf{h}\|^2\|\textbf{g}\|^2\right)\right]\right)\right)\nonumber\\
&=&\frac{T-\tau}{T}\log_2\left(1+\frac{c\tau}{T-\tau}\exp\left(2E\left[\ln\left(\|\textbf{h}\|^2\right)\right]\right)\right),\label{eqn16}
\end{eqnarray}
where $c=a\left(1-2^{-\frac{B}{N_t-1}}\right)$. (\ref{eqn14})
holds true because $\log_2(1+\exp(x))$ is convex in $x$ and the
desired results can be obtained by applying Jensen's inequality.
In (\ref{eqn15}), we replace
$|\tilde{\textbf{h}}^H\textbf{w}_{i^{\star}}|^2$ with
$1-2^{-\frac{B}{N_t-1}}$, which is numerically quite accurate.
(\ref{eqn16}) follows from the fact that $\textbf{h}$ and
$\textbf{g}$ are i.i.d.. It is worth noting that an approximation
is used in (\ref{eqn15}), so (\ref{eqn16}) is an approximate low
bound on the average information transmission rate. Intuitively,
the key of deriving the approximate lower bound is to compute the
expectation in (\ref{eqn16}). $\|\textbf{h}\|^2$ is $\chi^2$
distributed with $2N_t$ degrees of freedom, denoted by $g(x)$,
then $E\left[\ln\left(\|\textbf{h}\|^2\right)\right]$ can be
computed as
\begin{eqnarray}
E\left[\ln\left(\|\textbf{h}\|^2\right)\right]
\!\!\!\!&=&\!\!\!\int_0^{\infty}\ln(x)g(x)dx\label{eqn17}\\
\!\!\!\!&=&\!\!\!\frac{1}{\Gamma(N_t)}\int_0^{\infty}\ln(x)x^{N_t-1}\exp(-x)dx\label{eqn18}\\
\!\!\!\!&=&\!\!\!\psi(N_t),\label{eqn19}
\end{eqnarray}
where (\ref{eqn19}) is derived according to [Eq. 4.3521, 11], and
$\psi(z)=\frac{d\Gamma(z)}{dz}$ is the Digamma function function
and $\psi(N_t)=\sum\limits_{k=1}^{N_t-1}\frac{1}{k}-C$,
$C=0.57721566490$ is the Euler constant. Substituting
(\ref{eqn19}) into (\ref{eqn16}), we get an approximate lower
bound on the average information transmission rate as
\begin{equation}
R^{l}(\tau)=\frac{T-\tau}{T}\log_2\left(1+\exp\left(2\psi(N_t)\right)\frac{c\tau}{T-\tau}\right).\label{eqn20}
\end{equation}
Therefore, the wireless energy and information transfer based on
the approximate lower bound in (\ref{eqn20}) can be formulated as
the following optimization problem:
\begin{eqnarray}
J_2\!\!\!\!\!&:&\!\!\!\!\!
\max\limits_{\tau}\frac{T-\tau}{T}\log_2\left(1+\frac{d\tau}{T-\tau}\right)\label{eqn21}\\
&s.t.&0\leq\tau\leq T,\label{eqn22}
\end{eqnarray}
where $d=\eta\alpha\theta
P_1\left(1-2^{-\frac{B}{N_t-1}}\right)\exp\left(2\psi(N_t)\right)/\sigma^2$.
Notice that $J_2$ is similar to $J_1$, except the constant
coefficient $d$ instead of $b$. So the optimal energy transfer
duration can be derived according to the following criteria:
\begin{equation}
\tau^{\star}=\arg\max\limits_{\tau\in\Omega_2}R^{u}(\tau),\label{eqn23}
\end{equation}
where $\Omega_2=\{\tau_1,\tau_2,\tau_4\}$, and $\tau_4$ is the
solution of the following equation:
\begin{eqnarray}
\frac{c}{(T+(c-1)\tau)\ln2}-\frac{1}{T}\log_2\left(1+\frac{c\tau}{T-\tau}\right)=0,\label{eqn24}
\end{eqnarray}
which is derived by letting the derivative of the objective
function in (\ref{eqn21}) with respect to $\tau$ equal to zero.

\section{The Impact of Imperfect CSI}
In the last section, we derived the wireless energy and
information transfer tradeoffs in limited feedback multi-antenna
systems from the perspectives of the maximization of an upper
bound and an approximate lower bound on the average information
transmission rate respectively in the case of perfect CSI at
T$_2$. However, in practical systems, due to limited channel
estimation capacity, there is more or less estimation error at
T$_2$. A commonly used CSI error model is given by \cite{CEE}
\begin{equation}
\textbf{h}=\rho\textbf{h}_{e}+\sqrt{1-\rho^2}\textbf{n}_e,\label{eqn25}
\end{equation}
where $\textbf{h}$ and $\textbf{h}_{e}$ are the real and estimated
CSI, respectively. $\textbf{n}_e$ is the estimation error noise
distributed as $\mathcal{CN}(0, \textbf{I}_{N_t})$, and is
independent of $\textbf{h}_{e}$. $\rho$ is the correlation
coefficient between $\textbf{h}$ and $\textbf{h}_{e}$, and larger
$\rho$ means higher estimation precision. As seen in (\ref{eqn1}),
the energy beamforming vector is selected based on the estimated
CSI, so CSI mismatch may result in SNR loss. Moreover, we assume
T$_1$ has the perfect CSI of $\textbf{g}$. Under this condition,
the harvesting power and the average information transmission rate
are given by
\begin{eqnarray}
Q_{harv}^{'}&=&\eta\alpha P_1|\textbf{h}^H\textbf{w}_{i^{\star}}|^2\nonumber\\
&=&\eta\alpha
P_1(\rho^2|\textbf{h}_e^H\textbf{w}_{i^{\star}}|^2+(1-\rho^2)),\label{eqn26}
\end{eqnarray}
and
\begin{eqnarray}
R_e(\tau)&=&\frac{T-\tau}{T}E\bigg[\log_2\bigg(1+\frac{a\rho^2\|\textbf{h}_e\|^2|\tilde{\textbf{h}_e}^H\textbf{w}_{i^{\star}}|^2\|\textbf{g}\|^2\tau}{T-\tau}\nonumber\\
&+&\frac{a(1-\rho^2)|\textbf{n}_e^H\textbf{w}_{i^{\star}}|^2\|\textbf{g}\|^2\tau}{T-\tau}\nonumber\\
&+&\frac{2a\rho\sqrt{1-\rho^2}\mathcal{R}(\textbf{w}_{i^{\star}}^H\textbf{h}_e\textbf{n}_e\textbf{w}_{i^{\star}})\|\textbf{g}\|^2\tau}{T-\tau}\bigg)\bigg],\label{eqn27}
\end{eqnarray}
respectively, $\mathcal{R}(x)$ denotes the real part of complex
value $x$. Thus, we can derive an upper bound on the average
information transmission rate in presence of imperfect CSI as
\begin{eqnarray}
R_e^{u}(\tau)&=&\frac{T-\tau}{T}\log_2\bigg(1+\frac{4a\rho^2
N_t^2\left(1-\frac{N_t-1}{N_t}2^{-\frac{B}{N_t-1}}\right)\tau}{T-\tau}\nonumber\\
&+&\frac{2a(1-\rho^2)N_t\tau}{T-\tau}\bigg).\label{eqn28}
\end{eqnarray}
Compare (\ref{eqn28}) with (\ref{eqn9}), it is found that the
impact of imperfect CSI at T$_2$ is to replace the term
$4aN_t^2\left(1-\frac{N_t-1}{N_t}2^{-\frac{B}{N_t-1}}\right)\frac{\tau}{T-\tau}$
with $\left(4a\rho^2
N_t^2\left(1-\frac{N_t-1}{N_t}2^{-\frac{B}{N_t-1}}\right)+2a(1-\rho^2)N_t\right)\frac{\tau}{T-\tau}$.
Notice that only when $4a\rho^2
N_t^2\left(1-\frac{N_t-1}{N_t}2^{-\frac{B}{N_t-1}}\right)+2a(1-\rho^2)N_t>4aN_t^2\left(1-\frac{N_t-1}{N_t}2^{-\frac{B}{N_t-1}}\right)$,
namely $B<-(N_t-1)\log_2\left(\frac{2N_t-1}{2(N_t-1)}\right)<0$,
the performance in presence of imperfect CSI is better than that
under ideal condition based on the same wireless energy and
information transfer tradeoff. However, $B$ is impossible to be
smaller than zero, so imperfect CSI will lead to the performance
loss inevitably. Furthermore, replacing $b$ in the optimization
problem $J_1$ with $a\rho^2
N_t^2\left(1-\frac{N_t-1}{N_t}2^{-\frac{B}{N_t-1}}\right)+a(1-\rho^2)N_t$,
we can derive the corresponding optimal tradeoff.

\section{Numerical Results}
To examine the effectiveness of the proposed algorithms for the
wireless energy and information transfer tradeoff in limited
feedback multi-antenna systems with energy beamfoming, we present
several numerical results in different scenarios. For convenience,
we set $N_t=4$, $\eta=0.8$, $\sigma^2=-125$dBm, $T=5$ms and
$\alpha=\theta=10^{-2}d^{-\nu}$ for all simulation scenarios,
where $d=10$m is the distance between T$_1$ and T$_2$ and $\nu=4$
is the path loss exponent. Note that in the above path loss model,
a 20dB path loss is assumed at a reference distance of 1m. In the
following, we use UA and LA to denote the proposed algorithms for
the wireless energy and information tradeoff based on an upper
bound and an approximate lower bound of the average information
transmission rate, respectively. For comparison, we use EA and OA
to denote the fixed equal duration allocation algorithm, namely
$\tau=T/2=2.5$ms, and the optimal algorithm by numerically
searching the optimal $\tau$ in (\ref{eqn5}), respectively.

\begin{figure}[h] \centering
\includegraphics [width=0.5\textwidth] {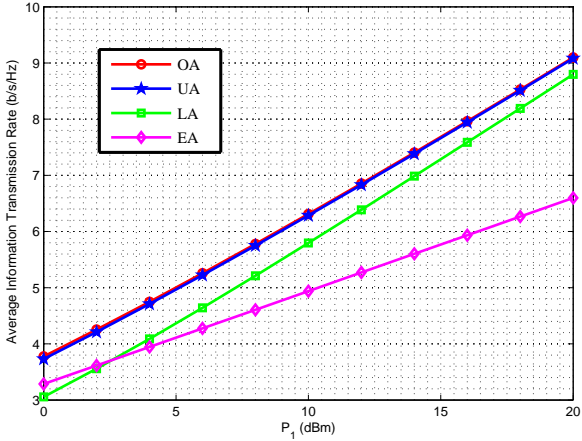}
\caption {Performance comparison of different tradeoff algorithms}
\label{Fig3}
\end{figure}

Fig.{\ref{Fig3}} compares the average information transmission
rates $R$ of the four wireless energy and information tradeoff
algorithms, namely UA, LA, EA and OA, when $B=4$. It is found that
the proposed UA can nearly achieve the same performance as the OA
in the whole transmit power region. Although the EA has the low
complexity, the proposed UA and LA perform better than the EA in
the moderate and high transmit power regions. Especially for the
UA, it provides more than 5dBm gains at $R=5$ b/s/Hz with respect
to the EA. As the transmit power increases, the performance gain
becomes larger. Therefore, the proposed UA is more appealing in
the practical systems from the perspectives of both the
performance and the complexity.

\begin{figure}[h] \centering
\includegraphics [width=0.5\textwidth] {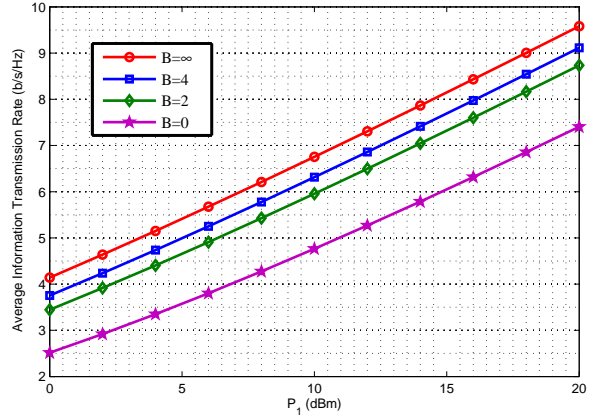}
\caption {Performance comparison of UA algorithm with different
codebook sizes} \label{Fig4}
\end{figure}

Next, we show the benefit of the proposed energy beamforming from
the perspective of feedback amount. In this scenario, we only use
the proposed UA due to its good performance as seen in
Fig.\ref{Fig3}. It is known that the larger feedback means more
CSI at T$_1$, so the efficiency of wireless energy transfer or
energy beamforming is higher due to transmission direction
alignment, and thus the average information transmission rate is
larger. As seen in Fig.\ref{Fig4}, compared with no energy
beamforming ($B=0$), energy beamforming even with a very small
codebook ($B=2$) can obtain an obvious performance gain. For
example, at $P_1=1$dB, there is more than 1 b/s/Hz gain. Moreover,
with the increase of transmit power, the performance gain becomes
larger. On the other hand, for energy beamforming, the performance
gain by adding from $B=2$ to 4, and from $B=4$ to that with full
CSI, namely $B=\infty$, is not as large as $B=0$ to $B=2$.
Nonetheless, the CSI feedback plays an important role and we can
control the feedback amount or the codebook size to balance the
performance and feedback overhead.

\begin{figure}[h] \centering
\includegraphics [width=0.5\textwidth] {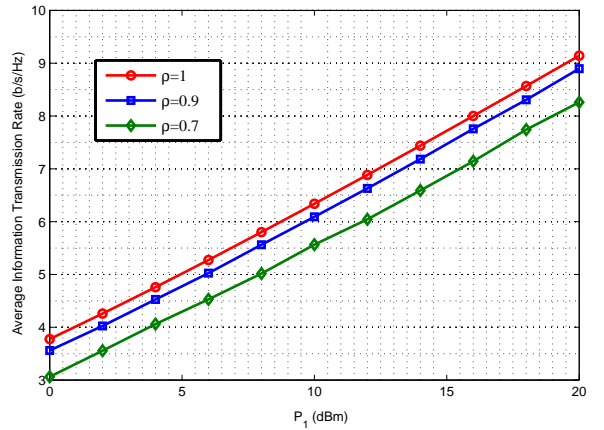}
\caption {Performance comparison of UA algorithm with different
correlation coefficients} \label{Fig5}
\end{figure}

Finally, we investigate the impact of imperfect CSI at T$_2$ on
the average information transmission rate based on the proposed UA
with $B=4$. As analyzed earlier, we use correlation coefficient
$\rho$ to denote the precision of CSI at T$_2$. Smaller $\rho$
means severe CSI mismatch. It is shown in Fig.\ref{Fig5} that when
CSI estimation is relatively accurate, i.e $\rho=0.9$, the
performance loss is very small. With the decrease of $\rho$, there
is a gap with respect to the case of $\rho=1$, namely perfect CSI
at T$_2$.

\section{Conclusion}
A major contribution of this paper is the introduction of the
wireless energy and information transfer tradeoff, so as to
maximize the average information transmission rate. Moreover, we
also propose to perform energy beamforming in the multiantenna
wireless energy and information transfer system to optimize the
efficiency of energy transfer, via the CSI feedback based on a
quantization codebook. For a given codebook size, we derive an
upper bound and an approximate lower bound of the average
information transmission rate. By maximizing the two bounds, we
obtain two tradeoff algorithms. Furthermore, we investigate the
impact of imperfect CSI on the average information transmission
rate, and present the corresponding performance.

\section*{Acknowledgment}
We thank the reviewers and the associate editor for the careful
reviews and for their suggestions which helped in improving the
quality of the paper.

\end{document}